
{}.
\vskip 1.5true in
\centerline{TOTAL AND JET PHOTOPRODUCTION CROSS SECTIONS AT HERA}
\vskip 0.5true in
\centerline{Ina Sarcevic}
\centerline{Physics Department}
\centerline{University of Arizona}
\centerline{Tucson, AZ 85721}
\vskip 3.0true in
\noindent
{ABSTRACT \hfill}
\vskip 0.2true in
\noindent
We show that the recent measurements of the total photoproduction cross
section at HERA energies [1] are in agreement with our earlier
prediction based on
high-energy
hadronic structure of the photon in
the QCD minijet-type  model [2].  We discuss
an improved calculation of the photoproduction cross section,
in which the soft part,
motivated by the Regge theory,
is taken to be energy-dependent
and the semi-hard hadronic part is carefully eikonalized
to take into account for the multiple scatterings and to
include the appropriate
hadronic probability of the
photon [3].  We show
that the extrapolation of our cross section
to ultra-high energies, of relevance
to the cosmic ray physics, gives significant contribution
to the ``conventional'' value,
but cannot account for the anomalous muon content
observed in the cosmic ray air-showers associated with
astrophysical point
sources [4].
\vfil\eject
\magnification=1100
\baselineskip=16pt
\nopagenumbers
\hsize 17true cm
\vsize 25 true cm
\voffset -1.1 true cm
\hoffset -0.3 true cm

\noindent
1.  INTRODUCTION
\vskip 0.041true in

One of the most striking aspects of high energy
photoproduction
is the unusual hadronic character of the
photon, the fact that photon can produce
$q \bar q$ pair, and then through
 subsequent QCD evolution fill up the confinement volume with
quarks and gluons with a density akin to that of a pion or
nucleon [2,3].
The probability of photon acting hadronicaly increases with
energy and therefore it is not surprising that
measurements of the total photoproduction cross section
up to $\sqrt s=18GeV$ [5]
show rise with energy
similar to the one
observed in hadronic collisions [6].  Recent HERA measurements [1]
make this even more pronounced.
In hadronic collisions, the rapid growth
of the total cross section
was associated with a
dominance of hard scattering partonic processes
over the nonperturbative (soft) ones [7], supported by the
detection
of the semi-hard QCD jets (so-called
``minijets'') at CERN Collider energies [8].
Similarly, recent
observation of the
hard scatterings in photoproduction at HERA energies
corroborates this
hypothesis [9].
\vskip 0.041true in
\par
\noindent
2.  PHOTOPRODUCTION CROSS SECTIONS
AT HERA ENERGIES
\vskip 0.041true in
\par
The total photoproduction cross
section measured in the energy range $10GeV\leq \sqrt s\leq 18GeV$
 and, most recently, at HERA energies points
towards
the hadronic behavior of the photon.
Few years ago,
we have made predictions for the total and jet
photoproduction cross sections
in a simple QCD minijet-type model based on analogy with
hadronic collisions [2].
We have assumed that
the total photoproduction cross section
can be represented as
a sum of the soft (nonperturbative) and hard (jet) part
(i.e. $\sigma_{T} = \sigma_{soft} + \sigma_{JET} $), where
the soft part is energy
independent, determined
from the low
energy
data ($\sqrt s\leq 10GeV$).
The jet (hard) part has contributions from
two subprocesses:
the ``standard'' (direct) QCD process ($\gamma q\rightarrow q g$ and
$\gamma g\rightarrow q\bar q$) and the ``anomalous'' process (for
example, $\gamma\rightarrow q\bar q$, followed by quark
bremsstrahlung,
$q\rightarrow qg$ and $gg\rightarrow gg$).
The later process is the same as the jet production process in
p-p collisions up to a photon structure function.
We note that the photon structure function is
proportional to
$\alpha_{em}/\alpha_{s}$, where $\alpha_{em}$ is the
 electromagnetic
coupling. The effective order of the above processes is therefore
  $ \alpha_{em}\alpha_{s},
$ since the jet cross sections are of order $\alpha_{s}^{2}$.
Thus, they
 are  of
the same order as direct  two-jet processes, in which the
photon-parton
vertex is electromagnetic and does not involve the photon's
hadronic content.
The
existing parametrizations of the photon structure function,
Duke and Owens (DO), Drees and Grassie (DG) and Abramowicz et
al. (LAC1) [10],
all describe
low energy photoproduction
data very well.  However, they differ dramatically at
very high energies (i.e. low $x$ region), the region of
the HERA experiment, for example.  Therefore
recent photoproduction
measurements at HERA energies can provide valuable information
about the photon structure function at
small $x$ ($x\sim 4p_t^2/s$), and in particular its gluon
content.
\par
The QCD jet cross section for
photon-proton interactions is given by
$$
\sigma_{\rm QCD}^{\gamma p}=
 \sum\limits_{ij}{1\over{1 + \delta_{ij}}}
  \int dx_{\gamma}dx_{p}
 \int_{p_{\perp,\rm {min}}^{2}}
  \hskip -20pt dp_{\perp}^2
  [f_{i}^{(\gamma)}(x_{\gamma},\hat{Q}^{2})f_{j}^{(p)}
  (x_{p},\hat{Q}^{2}) + i \leftrightarrow j]
 {d \hat{\sigma}_{ij} \over {dp_{\perp}^2}},
  \eqno(1) $$
\noindent where $\hat {\sigma_{ij}}$ are
parton cross sections and
$f_{i}^{(\gamma)} (x_{\gamma}, \hat{Q}^{2})$
$(f_{j}^{(p)}(x_{p},
\hat{Q}^{2}))$ is the photon
(proton) structure function.
The expressions for all the subprocesses that contribute to
$\hat {\sigma_{ij}}$ can be found in Ref. [11].
We take the choice
of scale $\hat {Q^2} = p_T^2$, which is shown to give very good
description of the hadronic jet data [11].
For parton
structure function we use EHLQ parametrization [12].
The results do not show appreciately sensitivity to
the choice of the proton structure function.

{}From the constant low energy data [5], we determine
the soft part of the cross section to be
$\sigma_{soft}=0.114mb$.  The observed $3\%$
increase of
the cross section in the energy range between $10GeV$ and $18GeV$
can be described by adding the
hard (jet) contribution
with jet transverse
momentum cutoff $1.4GeV\leq p_{\perp,\rm {min}}\leq 2GeV$
to the soft part [2].  The actual value of
$p_{\perp,\rm {min}}$, however, below which nonperturbative
processes make important contributions, is impossible to pin
down theoretically using perturbative techniques.
As the energy increases direct and soft part become
negligible in comparison with the anomalous part,
because the later has a steep increase with energy.
We find that
in the Fermilab E683 energy range ($\sqrt s\leq 28GeV$), the
results for the cross sections are not sensitive to the choice
of the photon structure function.  Therefore,
in addition to providing important confirmation of the
hadronlike nature of photon-proton interactions,
one could use
forthcoming E683 experiment to pin down the theoretical
parameter $p_{\perp,\rm {min}}$ to a few percent.
\par
In Fig. 1 we show our results for
$\sigma_{\gamma p}$ at HERA
energies [2].  We note that the results are very sensitive to the
choice of the photon structure function due to their
different $x$ behavior at very high energies. For example,
DO gluon structure function behaves as $f_g^{\gamma}
= {x^{-1.97}}$,
while DG has less singular behavior, $f_g^{\gamma}\sim
{x^{-1.4}}$
at the scale $Q^2=p_T^2=4GeV$.
The
cross sections obtained using DG photon
structure function are more realistic, since the extrapolation
of DO parametrization to small $x$ region give unphysically
singular behavior.
For this reason, all the
cross sections obtained using DO function should
be treated as {\it extreme} theoretical upper bounds.
In Fig. 1 we also present the results for the cross section when only
soft and direct part are included, indicating its very weak energy
dependence.  The rise of the total cross section is thus
mostly driven by the
``anomalous'' (hadronic) part of the cross section.  We note that
HERA measurement has some resolving power to distinguish
between different sets of photon structure function and therefore
determine presently unknown low $x$ behavior of its gluon part.
For example, the cross sections
obtained using DO structure functions are already excluded by
HERA data, while
theoretical result obtained using DG structure function and
$p_{\perp,\rm min}=2GeV$
is consistent with the data (see Fig. 1).
However,
one should keep in mind that all the
theoretical predictions presented in Fig. 1 do not take into
account the possibility of multiple scatterings which, as we
will show later, will result in reducing
the cross sections by $10\%$ for $p_
{\perp,\rm min}=2GeV$ and by about $30\%$ for
$p_{\perp,\rm min}=1.41GeV$.
\par
We have also calculated the total
jet cross sections at
Fermilab and HERA energies for jet $p_T$ triggers of
$3,4,5GeV$ and $5,10,15GeV$ respectively [2].  The energy
dependence of the total jet cross section is much steeper than of the
direct part of the jet cross section only.
This can be seen in Figs. 4 and
5 in Ref. [2].
Again, jet measurements at HERA energies can
distinguish between different photon structure functions,
but
in this case one is probing the photon structure function
at slightly higher values of $x$ than in the case of the
total
cross section.
\par
Recently, we have improved our calculation of the total photoproduction
cross section by incorporating the
possibility of multiple
parton collisions in a single photon-nucleon collision
by using an
eikonal treatment of the high-energy scattering process [3].
In our general formulation of the theory
the $\gamma p$
cross section is expressed as a sum over properly eikonalized cross
sections for the interaction
\vfil\eject
{}.
\vskip 10.0 true cm
\vbox{
\baselineskip=8pt
\noindent
Figure 1:
Total inelastic cross section ($\sigma_{soft}+\sigma_{JET}$)
predictions for HERA energies [2], compared to the recent ZEUS and H1
measurements [1].  The jet part includes contributions from
direct processes.  Also shown separately are contributions
of direct processes, added to the constant soft part (curve $5$).
}
\baselineskip=16pt
\vskip 0.8true cm
\noindent
of the virtual hadronic components of the
photon with the proton, with each cross section weighted by the
probability with which that component appears in the photon.  We have
developed a detailed
model which includes contributions from light
vector mesons and from excited virtual states described in a
quark-gluon basis.  The parton distribution functions which appear can
be related approximately
to those in the pion, while a weighted sum
gives
the distribution functions for the photon.
The ``soft" contributions
to the eikonal functions are parametrized in the form expected from
Regge theory to assure that the calculated high-energy cross
sections connect smoothly with the cross sections measured
at lower
energies.  The parton distribution functions which appear in the
``hard scattering" contributions to the eikonal
functions can be
related approximately to those in the pion.  This approach allows
us to use the (not-well-determined) pion distributions to predict
the high-energy behavior of the $\gamma p$
cross section, and also
gives an approximate prediction for the parton distributions in the
photon  in terms of these in the pion.  We have also
developed an alternative approach based directly on the photon
distribution functions, in accord with the analysis done in
Ref. [2].

In our QCD-based eikonal model the total
inelastic $\gamma p$ cross section is given by [3]
$$
\sigma_{\rm inel}^{\gamma p} = \sigma_{\rm dir} +
\lambda {\cal P}_\rho \int d^2 b \left( 1 - e^{-2{\rm Re}\,\chi_{\rho
p}}\right)
 + \sum_q e_q^2 {{\alpha_{\rm em}}\over {\pi}}
\int_{Q_0^2} {{dp_{\perp 0}^2}\over {p_{\perp 0}^2}}
\int d^2 b
\left( 1 - e^{-2{\rm Re}\,\chi_{q\bar qp}}\right),
\eqno(2)$$
where
$$
2{\rm Re}\,\chi_{\rho p}(b,s) = A_{\rho p}(b)
\left[ \sigma_{\rm soft}(s) + \sigma_{\rm QCD}^{\rho p}(s) \right],
\eqno(3)$$
and
$$
2{\rm Re}\,\chi_{q\bar qp} (b,s,p_{\perp 0}) = A_{q\bar qp}
(b,p_{\perp 0}) \left[ \sigma_{\rm soft}^{q\bar qp}
(s, p_{\perp 0}) + \sigma_{\rm QCD}^{q\bar qp}
(s,p_{\perp 0}) \right].
\eqno(4)$$
\noindent
$A_{\rho p}$ and
$A_{q\bar qp}$ are the overlapping spatial densities of the
corresponding systems
and $\sigma_{\rm QCD}^{q\bar qp}$ is given by the
analog of Eq. (1) with $f_i^\rho$ replaced by $f_i^{q\bar q}
\approx (Q_0/ p_{\perp 0} )^2 f_i^\rho$.  In Eq. (2), the
parameter $\lambda$ is the ``weight'' for the low-mass
vector meson contributions.  For example, $\lambda=4/3$ for
equal $\rho p$, $\omega p$ and $\phi p$ cross sections and
$\lambda=10/9$ for
complete suppression of the $\phi$ contribution [3].

The incident hadronic systems in a $\rho p$ collision can interact
inelastically through soft as well as hard processes.
The soft scattering is dominant
at low energies.  Motivated by
Regge theory,  we parametrize
$\sigma_{\rm soft}$  as energy-dependent, i.e.

$$\sigma_{\rm soft}^{\rho p} (s) \approx \sigma_0 + \sigma_1 (s -
m_p^2)^{-1/2} + \sigma_2(s-m_p^2)^{-1}. \eqno(5)$$
We take
$\sigma_{\rm QCD}^{\rho p}$ to be equivalent to $\sigma_{\rm
QCD}^{\gamma p}$ given by
Eq. (1).
We find
that the inelastic $\gamma p$ cross
section is strongly suppressed at high energies relative to
results reported earlier [13].
However, the QCD contributions to the hadronic interactions of
the photon still lead to a rapid rise in $\sigma^{\gamma p}$ at
HERA energies as predicted in earlier calculations [2] and
observed in recent experiments [1]. The magnitude of the rise provides
a quantitative test of the whole picture.
In particular, our results presented in Fig. 2,
show clearly that measurements of the
total inelastic $\gamma p$ cross section at HERA can impose strong
constraints on the parton distributions in the photon and, when
combined with low-energy measurements, it can pin down the
value of the theoretical cutoff
$p_{\perp,\rm min}$,
used to determine the onset of hard-scattering
processes.
{}From Fig. 2 we note that the cross sections obtained
using the value of the cutoff
$p_{\perp,\rm min}=2GeV$ seem to be a bit too
low for the observed increase
of the cross section in the energy range between $10GeV$ and $18GeV$,
but is in excellent agreement with ZEUS and H1 data,
while our results with
$p_{\perp,\rm min}=1.41GeV$ are in better agreement with the
data at all
energies when pion structure function is used and
slightly too large at HERA energies when DG structure function is
used for the photon.
Comparison of Fig. 1 and  Fig. 2 shows that
the eikonalization effect is to reduce the
cross sections
at HERA energies by
about $10\%$ for the case of
$p_{\perp,\rm min}=2GeV$ and
by almost
$30\%$ when
$p_{\perp,\rm min}=1.41GeV$ is used.
\vskip 7.2true cm
\vbox{
\baselineskip=8pt
\noindent
Figure 2:
The predictions for $\sigma_{\rm inel}^{\gamma p}$ for
$\sqrt{s}\leq400$ GeV for the QCD model based on pion structure
functions (solid and dashed curves)
and
on the Drees-Grassie
structure functions for the photon (dash-dotted and dotted curves)
with the $\phi$ meson contribution present or totally
suppressed and
with a)  $p_{\perp,\rm min}=
1.2$ GeV and $1.4 GeV$, b)
$p_{\perp,\rm min}=
2.0 $ GeV.
The lower-energy data are from [5] and the
HERA data at $\sqrt{s}\approx200$ GeV are from
[1].
}
\vfil\eject
\noindent
3.  THE HADRONIC PHOTON AND THE ``MUON PUZZLE''
\vskip 0.04true in
The unusually large cross sections at very high energies play
an
important role in
understanding recently observed anomalous muon
content in cosmic ray air-showers associated with
astrophysical
``point'' sources (such as Cygnus X-3, Hercules X-1 and Crab
Nebula) [4].  The number of muons
observed is
comparable with what one would expect in a hadronic shower, but
the fact that primary particle has to be long-lived and
neutral,
makes photon the only candidate in the Standard Model.
Conventionally, one would expect that photon produces electromagnetic
cascade and therefore muon poor.  However, if the photonuclear
cross section at very high energies becomes comparable with pair
production and bremsstrahlung
cross section ($\sigma_{\gamma \rightarrow
e^+e^-}\sim 500mb$) the muon content in a photon initiated
shower will be affected [13].  The hadronic character of the photon
enchances the photonuclear cross section at very high energies.
We have calculated the
total inelastic photon-air cross
sections in a QCD-based diffractive model, which takes into
account unitarity constraints necessary at
ultra-high energies [3].
\par
Our model for the hadronic interactions of the photon can be
extended to photon-nucleus interactions using Glauber's multiple
scattering theory with the result [3]
$$
\sigma_{\rm inel}^{\gamma-\rm air}
= A \sigma_{\rm dir} + \lambda{\cal P}_\rho
\int d^2 b \,\langle \Psi| 1 -
{\rm exp}\left(-\sum\nolimits_{j=1}^A2{\rm Re}\,\chi_j^{\rho p}
\right)
|\Psi\rangle+$$
$$ +\sum_q e_q^2 {{\alpha_{em}}\over {\pi}}
\int_{Q_0^2} {{dp_{\perp0}^2}\over {p_{\perp0}^2}}
\int d^2 b\,\langle \Psi |\,
1-{\rm exp}\left( -\sum\nolimits_{j=1}^A 2{\rm Re}\chi_j^{q\bar qp}
\right) |\Psi \rangle.
\eqno(6)$$
The expectation values are to be taken in the nuclear ground
state. ${\rm Re}\chi_j^m={\rm Re}\chi^m(|{\bf b}-{\bf r}_{j
\perp}|)$ is the eikonal function for the scattering of the
hadronic component $|m\rangle$ of the photon on the $j^{\rm th}$
nucleon, where  ${\bf r}_{j\perp}$ is the instantaneous
transverse distance of that nucleon from the nuclear center of
mass, and $\bf b$ is the impact parameter of the photon relative
to the nucleus. We have evaluated the integrals using shell-model
wave functions for the oxygen and nitrogen nuclei. The
eikonal functions for $\gamma n$ and $\gamma p$ scattering were
taken as equal.

Our results for the photon-proton
and photon-air cross
sections at energies $10\leq \sqrt s\leq 10^4$ are
presented in Fig. 6 in Ref. [3].
We find that $\sigma_{\gamma - p}$ at $\sqrt s\geq 10^3 GeV$ is
about four times
larger than the conventional value, large enough to
be interesting, but much too small to account for the reported
muon anomalies in photon-initiated showers.
\vskip 0.02true in
\noindent
4.  CONCLUSION
\vskip 0.02true in

We have shown how measurement of the total
photoproduction cross section
at
HERA energies provides valuable information about
the hadronic
character of the photon.
In particular we emphasized how
measurements of $\sigma_{\gamma p}$ at HERA energies
can impose strong constraint on the value
of the theoretical jet momentum cutoff and, more importantly,
determine the low $x$ behavior
of the photon structure function and its gluon content.
With the theoretical uncertainties being reduced, we have
extrapolated our predictions for the photonuclear cross section to
ultra-high energies relevant for cosmic ray experiments.
The new results on
$\gamma$-air interactions make it quite clear that the hadronic
interactions of the photon cannot explain the reported muon
anomalies in cosmic ray air-showers, if the anomalies in fact exist.
Furthermore, future cosmic
ray experiments might be able to
put the ``muon puzzle'' observations on
firmer grounds and to provide valuable input to
particle physics at ultra-high energies, currently far beyond
the range of accelerator
experiments.

\vskip 0.1true in
\baselineskip=14pt

\centerline{ACKNOWLEDGEMENTS}
\vskip 0.02true in
\par
The work presented here was done in collaboration with
R. Gandhi, L. Durand, K. Honjo and H. Pi whom I would like
to thank for many
useful discussions.  I am especially indebted
to K. Honjo for providing
Fig. 2 for this manuscript.
This work was
supported in
part by the United States Department
of Energy, Division of
High Energy and Nuclear Physics.
\vskip 0.1true in

\centerline{REFERENCES}
\vskip 0.02true in
\item{1.}  M. Derick {\it{et al.}},
ZEUS Collaboration, {\rm Phys. Lett.} {\bf B293}, 465 (1992);
T. Ahmed {\it et al.},
H1 Collaboration, {\rm Phys. Lett.} {\bf B299}, 374 (1993).
\item{2.} R. Gandhi and I. Sarcevic, {\rm Phys. Rev.}
{\bf D44}, 10 (1991).
\item{3.} L. Durand, K. Honjo, R. Gandhi, H. Pi and I. Sarcevic,
{\rm Phys. Rev.} {\bf D47}, 4815 (1993);
University of Wisconsin
preprint, MAD/TH/92-6, to be published in
{\rm Phys. Rev.} {\bf D}.
\item{4.} M. Samonski and W. Stamm, {\rm Ap. J.}~{\bf{L17}},
268 (1983);
B. L. Dingus {\it et. al.}, {\rm Phys. Rev. Lett}~
{\bf{61}}, 1906 (1988);
Sinha {\it et al.}, Tata Institute preprint, OG 4.6-23;
T. C. Weekes, {\rm Phys. Rep.}~{\bf{160}}, 1 (1988), and reference
therein.
\item{5.} D. O. Caldwell {\it et al.},
{\rm Phys. Rev. Lett.} {\bf 25},
609 (1970);
{\rm Phys. Rev. Lett.}
{\bf 40},
1222 (1978);
H. Meyer {\it et al.}, {\rm Phys. Lett.} {\bf 33B}, 189 (1970);
T. A. Armstrong {\it et al.}, {\rm Phys. Rev.}
{\bf D5}, 1640 (1972);
S. Michalowski {\it et al.}, {\rm Phys. Rev. Lett.}
{\bf 39}, 733 (1977).
\item{6.} N. Amos {\it et al.},
{\rm Nucl. Phys.} {\bf B262} 689 (1985); R. Castaldi and
G. Sanguinetti, {\rm Ann. Rev. Nucl. Part. Sci.}
{\bf 35}, 351 (1985);
 M. Bozzo {\it et al.},
{\rm Phys. Lett.} {\bf 147B}, 392 (1984); G. Alner {\it et al.},
 {\rm Z. Phys.} {\bf C32}, 156 (1986);
T. Hara {\it et al.}, {\rm Phys. Rev Lett.} {\bf 50}, 2058
 (1983); R. M. Baltrusaitis
 {\it et al.}, {\rm Phys. Rev. Lett.} {\bf 52}, 1380 (1984).
\item{7.} D. Cline, F. Halzen and J. Luthe, {\rm Phys. Rev. Lett.}
{\bf 31}, 491 (1973);
S. D. Ellis and M. B. Kislinger, {\rm Phys. Rev.}
{\bf D9}, 2027 (1974);
T. K. Gaisser and F. Halzen, {\rm Phys. Rev. Lett.}
{\bf 54}, 1754 (1985);
L. Durand and H. Pi, {\rm Phys. Rev. Lett.} {\bf 58}, 303 (1987);
{\rm Phys. Rev.} {\bf D38}, 78 (1988).
\item{8.} C. Albajar {\it{et al.}}, UA1 Collaboration,
{\rm Nucl. Phys.} {\bf B309}, 405 (1988).
\item{9.}  T. Ahmed {\it et al.}, H1 Collaboration,
{\rm Phys. Lett.} {\bf B297}, 205 (1992); M. Derick {\it et al.},
ZEUS Collaboration,
{\rm Phys. Lett.} {\bf B297}, 404 (1992).
\item{10.} M. Drees and K. Grassie, {\rm Z. Phys.}
{\bf C28}, 451 (1985);
D. Duke and J. Owens, {\rm Phys. Rev.}
{\bf D26}, 1600 (1982); H. Abramowicz
{\it et al.}, {\rm Phys. Lett.} {\bf B269}, 458 (1991).
\item{11.} I. Sarcevic, S. D. Ellis and P. Carruthers,
{\rm Phys. Rev.} {\bf D40}, 1472 (1989).
\item{12.} E. Eichten, I. Hinchliffe, K. Lane and C. Quigg,
{\rm Rev. Mod. Phys.} {\bf 56}, 579 (1984).
\item{13.} R. Gandhi, I. Sarcevic, A. Burrows, L. Durand and H. Pi,
{\rm Phys. Rev.} {\bf D42}, 263 (1990).
\vfil\eject
\end